\documentclass[twocolumn,preprintnumbers,prb,fleqn]{revtex4}
\usepackage{graphicx,amsfonts,amsbsy}
\usepackage[dvips]{epsfig,color}
\begin{document}
\title{Enhancement of spin stiffness with dilution in a ferromagnetic Kondo lattice model}
\author{Subrat Kumar Das}
\email{subrat@iitk.ac.in}
\author{Avinash Singh}
\email{avinas@iitk.ac.in}
\affiliation{Department of Physics, Indian Institute of Technology Kanpur - 208016}
\begin{abstract}
Carrier-induced ferromagnetism is investigated in a diluted ferromagnetic Kondo 
lattice model for several ordered impurity arrangements on square and cubic lattices,
allowing for quantitative comparison with different theoretical pictures.
The spin stiffness is found to be optimized with respect to hole doping concentration,
exchange interaction strength, as well as spin dilution due to a competition involving magnitude of carrier spin polarization and its oscillation length scale. 
The ferromagnetic transition temperature determined within the spin-fluctuation theory 
is in good agreement with experimental values for $\rm Ga_{1-x}Mn_xAs$.                  
\end{abstract}
\maketitle
\section{Introduction}
The recent discovery of ferromagnetism in diluted magnetic semiconductors (DMS) 
such as $\rm Ga_{1-x} Mn_x As$,\cite{Ohno2,Matsu,Pot,Edmonds} 
with transition temperature $T_c \simeq 110$ K for Mn concentration 
$x \simeq 5\%$,\cite{Matsu} 
and $\simeq 150$ K in films for $x$ in the range $6.7-8.5\%$,\cite{Pot,Edmonds} 
has led to intensive efforts to increase $T_c$ in view of potential technological applications. Tremendous interest has also been generated in the novel ferromagnetism exhibited by these systems in which magnetic interaction between localized spins is mediated by doped carriers.
The long-range oscillatory nature of the carrier-mediated spin couplings
results in a variety of interesting features of the ferromagnetic state,
such as significant sensitivity of spin stiffness and 
transition temperature $T_c$ on carrier concentration, 
competing antiferromagnetic interaction and noncollinear ordering,
spin-glass behaviour, 
spin clustering and disorder-induced localization etc.,
as recently reviewed.\cite{bhatt,Timm}

DMS such as $\rm Ga_{1-x}Mn_x As$ are mixed spin-fermion systems 
in which the $S=5/2$ Mn$^{++}$ impurities replace Ga$^{+++}$, 
thereby contributing a hole to the semiconductor valence band. 
However, large compensation due to As antisite defects reduces the hole density 
to nearly $10 \%$ of Mn concentration, which plays a key role in the stabilization 
of long-range ferromagnetic order,
and also provides a complimentary limit to Kondo systems.
The interplay between itinerant carriers in a partially filled band 
and the localized moments is conventionally 
studied within a diluted ferromagnetic Kondo lattice model (FKLM),
wherein $- J {\bf S}_I . {\mbox{\boldmath $\sigma$}}_I $ 
represents the exchange interaction between the magnetic impurity spin ${\bf S}_I$ 
and the electron spin ${\mbox{\boldmath $\sigma$}}_I$.

In order to obtain deeper understanding of the role of dilution 
in terms of the perturbation to carrier spin bands due to 
spatially varying impurity field 
as well as impurity-field-induced Zeeman splitting, 
which control the long-range oscillatory nature of the 
impurity-spin couplings through the carrier spin polarization $\chi^0({\bf q},\omega)$,
we will consider a diluted FKLM with {\em ordered} impurity arrangements in this paper. 
Use of {\bf k-}space representation allows for much larger lattices,
thus permitting a more refined study of the competing spin interactions 
with respect to carrier concentration, impurity separation, interaction strength, 
and wave vector.
More importantly, it provides for a quantitative comparison 
with different theoretical pictures, which we briefly review below.

Long range ferromagnetic interaction between the impurity spins is mediated,
in the mean-field (Zener model) picture,
\cite{Dietl1,Taka1,Jung1,Dietl2,Jung2,Dietl3}
by a uniform carrier spin polarization,
which is caused, in turn, by a uniform site-averaged impurity magnetic field
within the virtual crystal approximation (VCA).
In the weak-field limit ($xJS << \epsilon_{\rm F}$),
the carrier spin polarization is proportional to 
the Pauli susceptibility $\chi_{\rm P}$,
and the transition temperature ($T_c \sim xJ^2 \chi_{\rm P}$) 
is therefore proportional to the Mn concentration $x$, 
$J^2$, the carrier effective mass $m^*$,
and $N(\epsilon_{\rm F}) \sim p^{1/3}$, where $p$ is the hole concentration.

An alternative mechanism for the coupling between impurity spins 
involves the carrier-mediated Ruderman-Kittel-Kasuya-Yosida (RKKY) 
interaction.\cite{Matsu}
The local magnetic field ${\bf B}_J = J {\bf S}_J $ 
of an impurity spin at site $J$ polarizes the electrons locally, 
and the mobile band electrons spread this magnetic polarization in a 
characteristic manner: ${\bf m}_I = \chi^0 _{IJ} {\bf B}_J$,
where $\chi^0 _{IJ}$ represents the carrier spin polarization.
Another impurity spin $\vec{S}_I$ couples to this local electronic magnetization, resulting in an effective RKKY interaction $J^2 \chi^0 _{IJ} {\bf S}_I. {\bf S}_J$. 
Within a mean-field (MF) treatment of the resulting Heisenberg model,\cite{Dietl1}  
a similar behaviour for the transition temperature ($T_c \sim xJ^2 \chi_{\rm P}$) 
is obtained. 

Within the generalized RKKY theory involving the non-linear magnetic response,\cite{dms}
the carrier-induced spin couplings $J^2 \chi^0 _{IJ}(J)$ go through a maximum 
with respect to both carrier doping concentration $p$ 
and the fermion-spin interaction strength $J$. 
This optimization behaviour can be qualitatively understood in terms of a competition
between the increasing magnitude of the carrier spin polarization $\chi^0 _{IJ}(J)$
and the increasing rapidity of its oscillation,
which limits the growth of the spin couplings.
Similar behaviour was observed in the diluted Hubbard model\cite{Singh2,Pandey} 
for the effective spin coupling $U^2\chi^0 _{IJ} (U)$.

Long-wavelength magnon excitations provide a composite measure of the carrier-induced
spin couplings in the ferromagnetic state, with vansihing spin stiffness signalling
instability due to competing antiferromagnetic (AF) spin interactions.
Magnon excitations as function of electron density $n$ in the conduction band 
and the spin-fermion coupling $J$ have been studied within the 
concentrated FKLM having a magnetic impurity at every lattice site,
in the context of heavy fermion materials,\cite{Sig} ferromagnetic metals Gd, Tb, Dy, 
doped EuX\cite{Donath} and manganites.\cite{Furu,Wang,Yunoki,Vogt} 
Magnon dispersion has also been obtained in the context of DMS,
where the spatially varying impurity field has been treated exactly numerically,\cite{Berciu2}
within the VCA where a uniform impurity-induced Zeeman splitting 
of the carrier spin bands is assumed,\cite{Konig2}
and within the coherent potential approximation (CPA).\cite{Bouzerar,Nol2}
Magnon spectrum and transition temperature have also been obtained recently 
for $\rm Ga_{1-x} Mn_x As$ and $\rm Ga_{1-x} Mn_x N$
by means of an effective Heisenberg model, whose exchange parameters are obtained 
from first-principle calculations.\cite{hilbert_nolting}

\begin{figure}
\includegraphics[angle=0,width=.4\columnwidth]{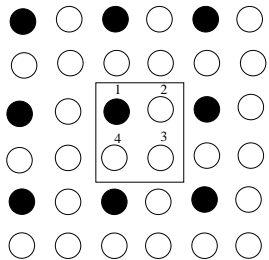}
\caption{An ordered arrangement of impurity spins ($\bullet$)
on a square host lattice ($\circ$).
Also shown are the sublattice labels corresponding to the four distinct
sites within the unit cell.}
\end{figure}

A sublattice-basis representation for ordered impurity arrangements is introduced 
in section II for two-dimensional (2d) and three dimensional (3d) cases,
and the formation of sub-bands due to spatially varying impurity field is discussed.
The two energy scales in the coupled spin-fermion problem are also introduced,
and temperature-dependence of impurity magnetization is obtained self-consistently.
Spin-wave excitations are obtained using the Holstein-Primakoff transformation 
in the large $S$ limit (section III).
A detailed study of the behaviour of spin stiffness with respect to 
dimensionality, carrier concentration, interaction strength as well as 
impurity separation (dilution) is presented in section IV.
Ferromagnetic transition temperature obtained within 
the renormalized spin-fluctuation theory is compared with 
RKKY and VCA results in section V. 

\section{Diluted Kondo lattice model}
We consider a diluted ferromagnetic Kondo lattice Hamiltonian
\begin{eqnarray}
H = t\sum_{i,\delta,\sigma} a_{i,\sigma}^\dagger a_{i+\delta,\sigma} -
\frac{J}{2} \sum_I{\bf S}_I.{\mbox{\boldmath $\sigma$}}_I
\label{h1}
\end{eqnarray}
on square and cubic lattices with positive hopping $t$ 
between nearest-neighbour (NN) sites $i,i+\delta$, which yields a host valence band. 
The second term represents the ferromagnetic exchange interaction 
between localized impurity spins at site $I$ and the fermion spin density 
${\mbox{\boldmath $\sigma$}}_I=\Psi_I^\dagger[{\mbox{\boldmath $\sigma$}}]\Psi_I$,
where the fermionic field operator 
$\Psi_I= \left ( \begin{array}{c} a_{I \uparrow} \\ a_{I \downarrow} 
\end{array} \right )$. The host on-site energy is set to zero. 

\subsection{Sublattice-basis representation}
We consider ordered (superlattice) arrangements of magnetic impurities
on square and cubic host lattices with impurity spacings $2a$ and $3a$.
Translational symmetry in the sublattice basis convenientally 
allows Fourier transformation to momentum space. 
For concreteness, we consider a square host lattice in the following, 
with magnetic impurities placed at every other host site,
corresponding to superlattice spacing $2a$ and impurity concentration $x=25\%$.
There are four sublattices, numbered $\alpha=1,2,3,4$, corresponding to the four
sites in the unit cell, as shown in Fig. 1.
We choose length and energy units such that the lattice spacing $a=1$ and 
the hopping term $t=1$.

In the MF approximation, the interaction term 
\begin{eqnarray}
H^{\rm int}_{MF}=- \frac{J}{2}\sum_I 
{\bf S}_I . \langle {\mbox{\boldmath $\sigma$}}_I \rangle  +
{\mbox{\boldmath $\sigma$}}_I . \langle {\bf S}_I \rangle 
\end{eqnarray}
represents a magnetic coupling of impurity and fermion spins with 
the self-consistently determined fermion and impurity fields 
$\frac{J}{2}\langle {\mbox{\boldmath $\sigma$}}_I \rangle$
and $\frac{J}{2}\langle {\bf S}_I \rangle \equiv {\bf B}_I$, respectively.
Assuming, without loss of generality,
a uniform impurity field ${\bf B}_I$ = $B\hat{z}$,
Fourier transformation within the sublattice basis 
yields the following MF Hamiltonian for spin-$\sigma$ fermion
\begin{equation}
H_{\rm MF}^{\sigma} ({\bf k}) =  \sum_{\bf k}\Psi_{{\bf k}\sigma}^\dagger \left [
\begin{array}{rccc}
-\sigma B & \epsilon_{\bf k}^{x} & 0 & \epsilon_{\bf k}^{y}  \\
\epsilon_{\bf k}^{x} & 0 & \epsilon_{\bf k}^{y} & 0  \\
0 & \epsilon_{\bf k}^{y} & 0 & \epsilon_{\bf k}^{x} \\
\epsilon_{\bf k}^{y} & 0 & \epsilon_{\bf k}^{x} & 0 \end{array} \right ] 
\Psi_{{\bf k}\sigma}
\end{equation}
where $\epsilon_{\bf k}^{x(y)} = 
2t\cos k_{x(y)}$ is the hopping energy along the $x(y)$ direction
and $\sigma=$+1 and $-1$ for spin up and down, respectively.
Here the field operator
$\Psi_{\bf k}= (a_{\bf k}^1 \; a_{\bf k}^2 \; a_{\bf k}^3 \; a_{\bf k}^4)$
defines the sublattice basis,
where $a_{\bf k}^\alpha $ refer to the fermion operator for sublattice
index $\alpha$.
The MF Hamiltonian is numerically diagonalized
to obtain the four eigenvalues $E_{{\bf k}\mu}^\sigma$,
corresponding to the four sub-band indices $\mu=1,2,3,4$,
and the four-component eigenvectors $\phi_{{\bf k}\mu}^{\sigma\alpha}$ yield the
fermion amplitudes on the sublattice $\alpha$.

\subsection{Quasiparticle density of states}
The quasiparticle density of states (Fig. 2) shows the 
formation of four and eight sub-bands for the square and cubic lattices,
respectively, corresponding to the number of sublattices.
The reflection symmetry about the origin of the energy axis 
reflects the magnetic coupling. 
The sub-band corresponding to impurity sublattice splits off and forms 
an impurity band at energy $\sim \pm B$,
which narrows with increasing $B$
due to decreasing effective hopping $t_{\rm eff} \sim t^2/B$ 
between impurity sites.
In the undoped case, both bands are completely filled,
and carrier doping introduces holes at top of the spin-$\downarrow$ impurity band.
With further hole doping, Fermi energy moves into the spin-$\uparrow$ band,
particularly for small $B$.

\begin{figure}
\begin{center}
\vspace*{-70mm}
\hspace*{-38mm}
\psfig{figure=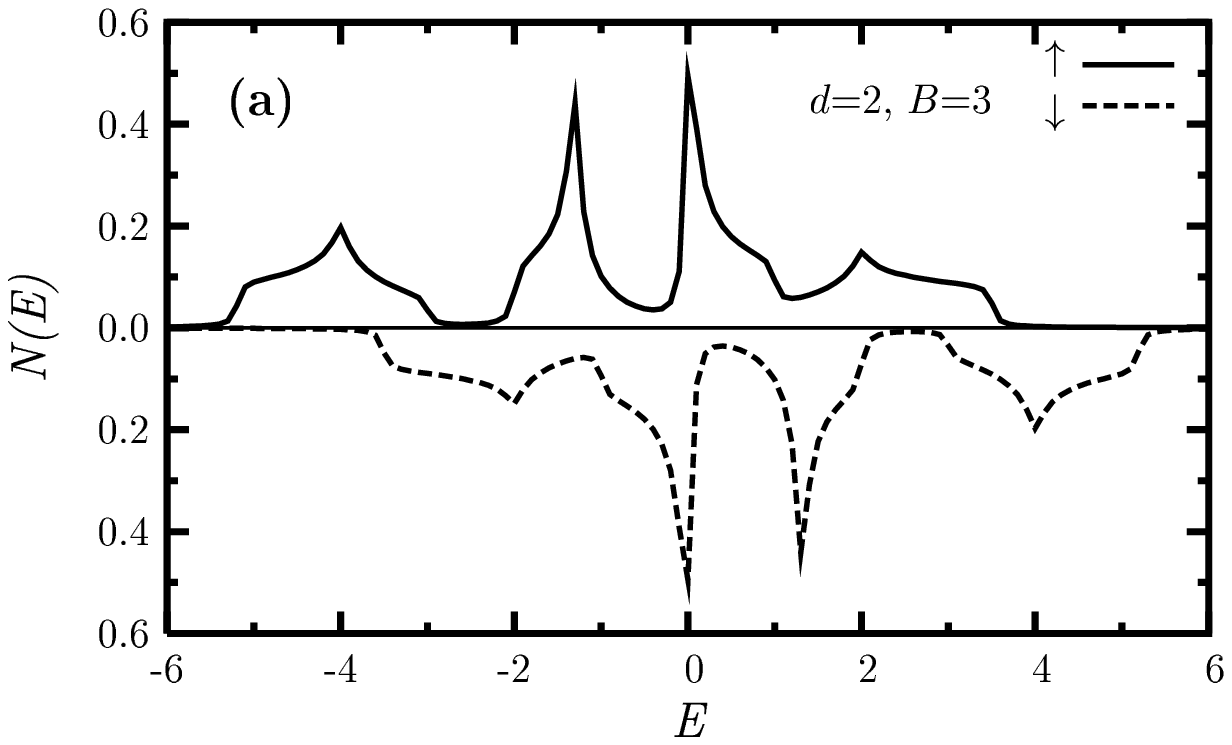,width=140mm}
\vspace{-77mm}
\vspace*{-75mm}
\hspace*{-38mm}
\psfig{figure=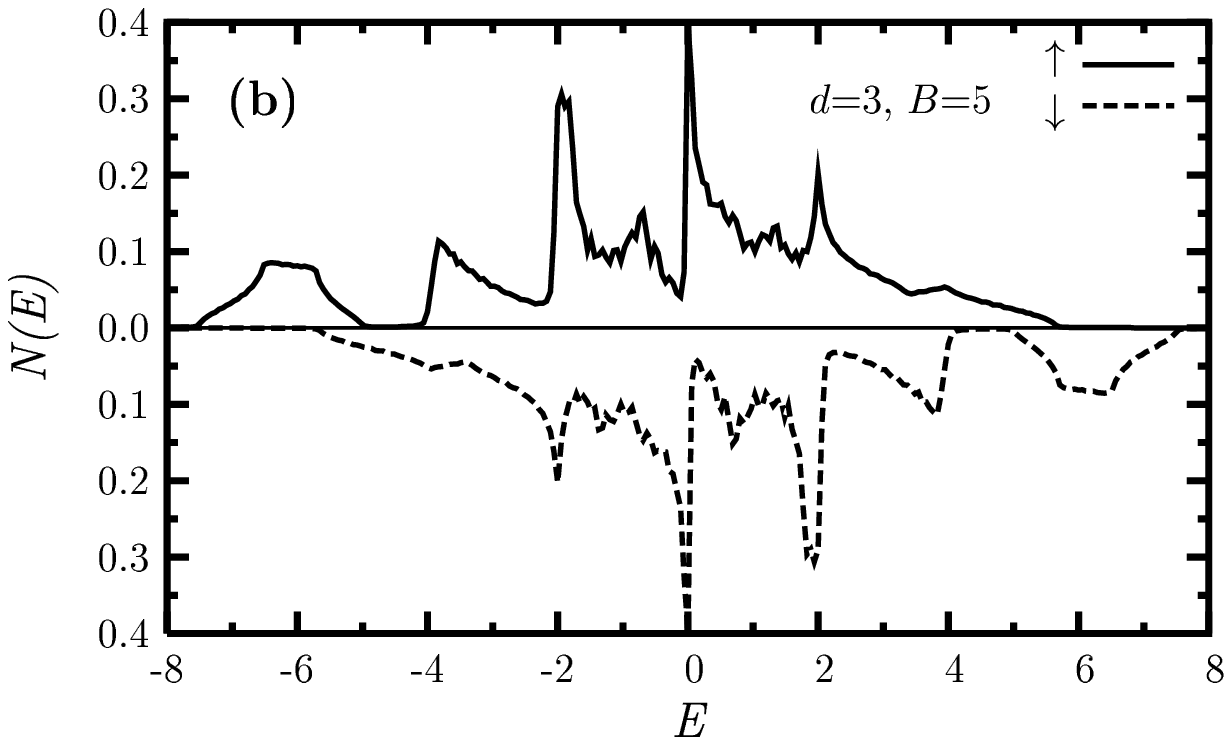,width=140mm}
\vspace{-85mm}
\end{center}
\caption{Spin-resolved density of states (DOS), showing the formation of sub-bands 
and impurity bands due to spatially-varying impurity field,
for 
(a) $d=2$, $x=25\%$ and
(b) $d=3$, $x=12.5\%$. The impurity field $B=JS/2$.}
\end{figure}

\subsection{Impurity magnetization}
With only one energy scale, the NN Heisenberg ferromagnet yields a transition temperature 
which differs from the MF value only by an O(1) number.
However, with long-range ferromagnetic order originating from the carrier-induced,
oscillating impurity-spin couplings, the transition temperature for DMS 
is very different from the MF value, which is determined from the bare spin-fermion
exchange interaction $J$, and physically represents a moment-melting temperature
rather than a spin-disordering temperature. 
In order to highlight the order of magnitude difference arising from 
competing interactions and spin softening, 
we have also obtained $T_c ^{\rm MF}$ in the MF approximation,
which is interesting in its own right, 
and studied recently for the $x=1$ case.\cite{Das2}

In the MF approximation, 
the coupled spin-fermion problem is self-consistently solved from the 
finite-temperature expectation values
\begin{eqnarray}
\langle \sigma_I^z \rangle & = & \sum_{{\bf k},\mu}
[(\phi_{{\bf k}\mu}^{\uparrow \alpha=1})^2-
(\phi_{{\bf k}\mu}^{\downarrow \alpha=1})^2] f_{\rm FD}(E_{{\bf k}\mu}^\sigma)  \\
\langle S_I^z \rangle & = & \sum_{m=-S} ^{+S} m e^{\beta mJ \langle \sigma_I^z \rangle /2} / 
\sum_{m=-S} ^{+S}  e^{\beta mJ \langle \sigma_I^z \rangle /2}
\end{eqnarray} 
of fermion and impurity spins, where $f_{\rm FD}$ is the Fermi-Dirac distribution function
and $\beta \equiv 1/k_{\rm B} T$. 

Consider the $x=1$ case, for simplicity. 
There are two distinct temperature scales associated with the fermion field 
$\frac{J}{2}\langle \sigma_I ^z \rangle \sim \frac{J}{2} p$ and the impurity field 
$\frac{J}{2}\langle S_I ^z \rangle \sim \frac{J}{2}S$, 
resulting in a characteristic concave
$\langle S_I ^z \rangle$ vs. $T$ behaviour  in the intermediate temperature regime
$Jp \ll k_{\rm B}T \ll JS$, wherein the nearly free impurity spins yield a typical 
paramagnetic $(\sim 1/T)$ response to impurity magnetization.  
In the diluted case, minority-spin holes preferentially sit on impurity sites,
resulting in an enhanced fermion magnetization $\langle \sigma_I ^z \rangle \sim p/x$,
which blurs the distinction between the two temperature scales.
Consequently, the concave behaviour in the temperature dependence of impurity 
magnetization [Fig. 3] becomes prominent only at low doping concentration $p$
and large $J$. 
With increasing temperature and decreasing impurity field, 
the rapid decrease in  $\langle \sigma_I ^z \rangle$
due to redistribution of holes from impurity to host sites 
results in a first-order transition, in contrast to the continuous transition 
obtained for the $x=1$ case.

\begin{figure}
\begin{center}
\vspace*{-70mm}
\hspace*{-38mm}
\psfig{figure=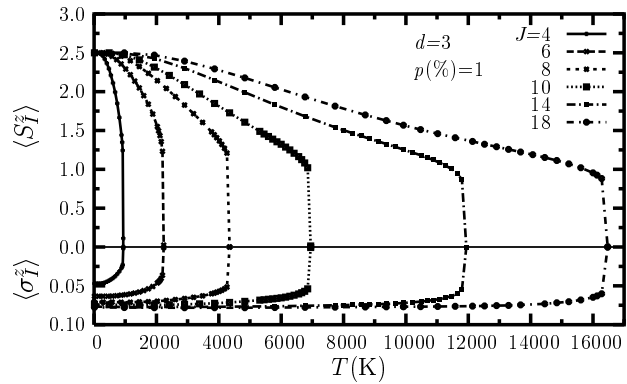,width=140mm}
\vspace{-85mm}
\end{center}
\caption{Variation of impurity and carrier magnetization with temperature,
showing a first-order transition for the 3d diluted case ($x=12.5\%$). 
Here host bandwidth $W=12t$ = 10 eV.} 
\end{figure}

\section{Spin-wave excitations}
Transverse spin fluctuations are gapless, low-energy excitations in the broken-symmetry
state of magnetic systems possessing continuous spin-rotational symmetry. 
Therefore, at low temperature they play an important role in diverse macroscopic 
properties such as existence of long-range order, magnitude and temperature dependence 
of the order parameter, magnetic transition temperature, spin correlations etc. 
In the following we consider $T=0$ and obtain the spin-wave excitations at the 
non-interacting level.
Finite impurity concentration, interaction strength $J$, and dynamical effects
are treated exactly in this approach.

Applying the approximate Holstein-Primakoff transformation in the large $S$ limit
\begin{eqnarray}
S_I ^+ &=&  b_I \sqrt{2S}  \nonumber \\
S_I ^- &=&  b_I ^\dagger \sqrt{2S} \nonumber \\
S_I ^z &=& S - b_I ^\dagger b_I 
\end{eqnarray}
from the spin-lowering ($S_I ^-$) and spin-raising ($S_I^+$) operators 
to boson (magnon) creation and annihilation operators
$b_I^\dagger$ and $b_I$,
which neglects magnon interaction terms of order $1/2S$,
the Kondo-lattice Hamiltonian reduces to  
\begin{eqnarray}
H &=& t\sum_{i,\delta,\sigma} a_{i,\sigma}^\dagger a_{i+\delta,\sigma} \nonumber \\
&-& \frac{J}{2} \sum_I \left [ \frac{\sqrt{2S}}{2} 
\left (b_I\sigma_I^- + b_I^\dagger \sigma_I^+ \right )
+ \left ( S -  b_I^\dagger b_I \right ) \sigma_I^z \right ] \; ,
\end{eqnarray}
where $\sigma_I^\pm \equiv \sigma_I ^x \pm i \sigma_I ^y$.

In the MF approximation, 
$\langle b_I^\dagger \rangle = \langle b_I \rangle = 
\langle b_I^\dagger b_I \rangle = 0$, 
and the Hamiltonian decouples into a fermion part
\begin{equation}
{\cal H}^0 _{\rm fermion} = t\sum_{i,\delta,\sigma} a_{i,\sigma}^\dagger 
a_{i+\delta,\sigma}
- \frac{J S}{2} \sum_I \sigma_I^z 
\end{equation}
and a boson part  
\begin{equation}
{\cal H}^0 _{\rm boson} = \frac{J}{2}\sum_I \langle \sigma_I^z \rangle  b_I^\dagger b_I
\; .
\end{equation}

We now obtain the time-ordered, transverse spin-fluctuation (magnon) propagator
\begin{eqnarray}
\chi_{IJ}^{+-}(t-t')=i\langle \Psi_{\rm G} \mid T[b_I (t) b_J ^\dagger (t')]\mid 
\Psi_{\rm G} \rangle
\end{eqnarray}
within the random phase approximation (RPA), by summing over all bubble diagrams 

\begin{figure}[hbt]
\input{chipm1.pstex_t}
\end{figure}
\hspace*{-5mm} 
where the particle-hole bubble in ${\bf q},\omega$ space 
\begin{eqnarray}
\chi^0({\bf q},\omega) &=& i \int \frac{d\omega'}{2\pi}G^{\uparrow}
({\bf k},\omega')G^{\downarrow}({\bf k-q},\omega'-\omega) \nonumber \\ 
&=& \sum_{E_{{\bf k}\nu }^\uparrow <E_F}^{E_{{\bf k-q} \mu }^\downarrow >E_F}
\frac{{|\phi_{{\bf k} \nu }^{\uparrow \alpha=1}|^2}
{|\phi_{{\bf k-q} \mu }^{\downarrow \alpha=1}|^2}}{E_{{\bf k-q} \mu }^\downarrow
- E_{{\bf k} \nu }^\uparrow + \omega}  \nonumber \\
&+& \sum_{E_{{\bf k} \nu }^\uparrow >E_F}^{E_{{\bf k-q} \mu}^\downarrow <E_F}
\frac{{|\phi_{k \nu}^{\uparrow\alpha=1}|^2}
{|\phi_{{\bf k-q} \mu}^{\downarrow \alpha=1}|^2}}{E_{{\bf k} \nu}^\uparrow 
- E_{{\bf k-q} \mu}^\downarrow - \omega}
\end{eqnarray}
involves integrating out the fermions in the broken-symmetry state.
It is the particle-hole bubble $\chi^0({\bf q},\omega)$ 
which mediates the carrier-induced impurity spin couplings in the ferromagnetic state, 
and the oscillatory, long-range nature of the spin couplings is 
effectively controlled by the Fermi wavevector ($k_F \sim p^{1/d})$ 
and the impurity-field-induced Zeeman splitting of the carrier spin bands.

\begin{figure}
\begin{center}
\vspace*{-70mm}
\hspace*{-38mm}
\psfig{figure=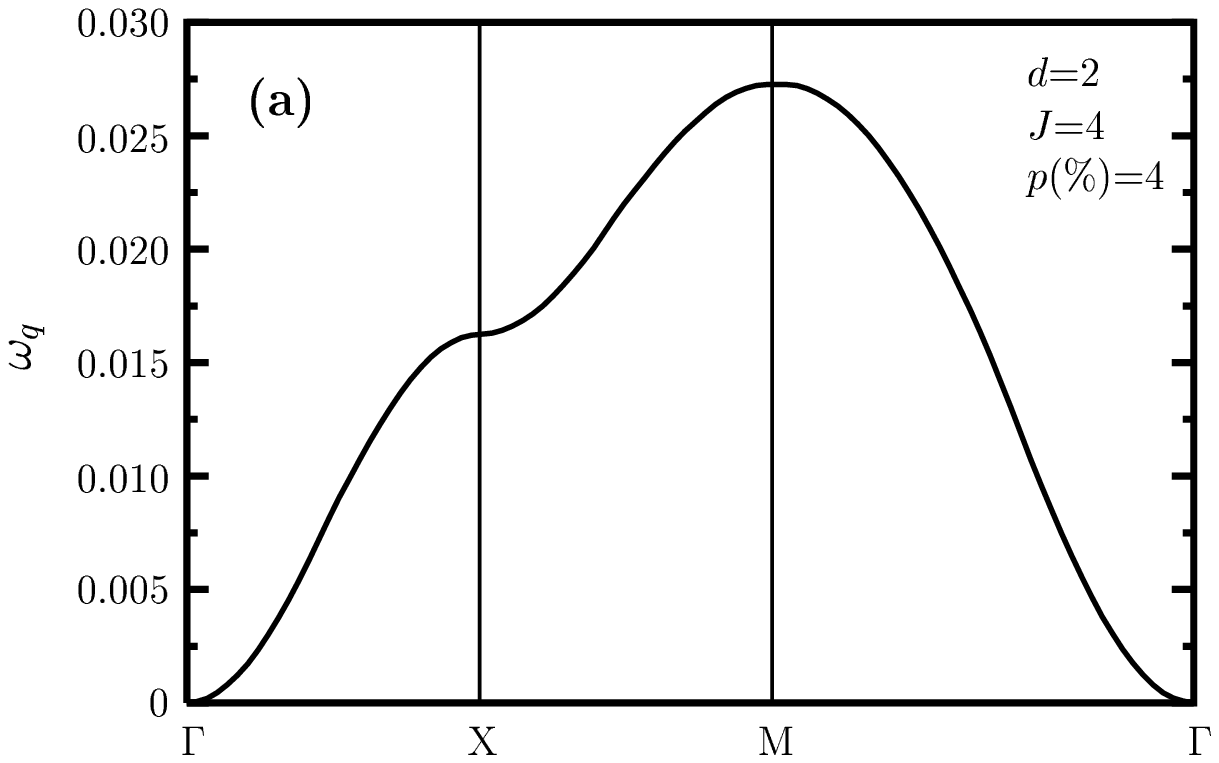,width=140mm}
\vspace{-77mm}
\vspace*{-74mm}
\hspace*{-38mm}
\psfig{figure=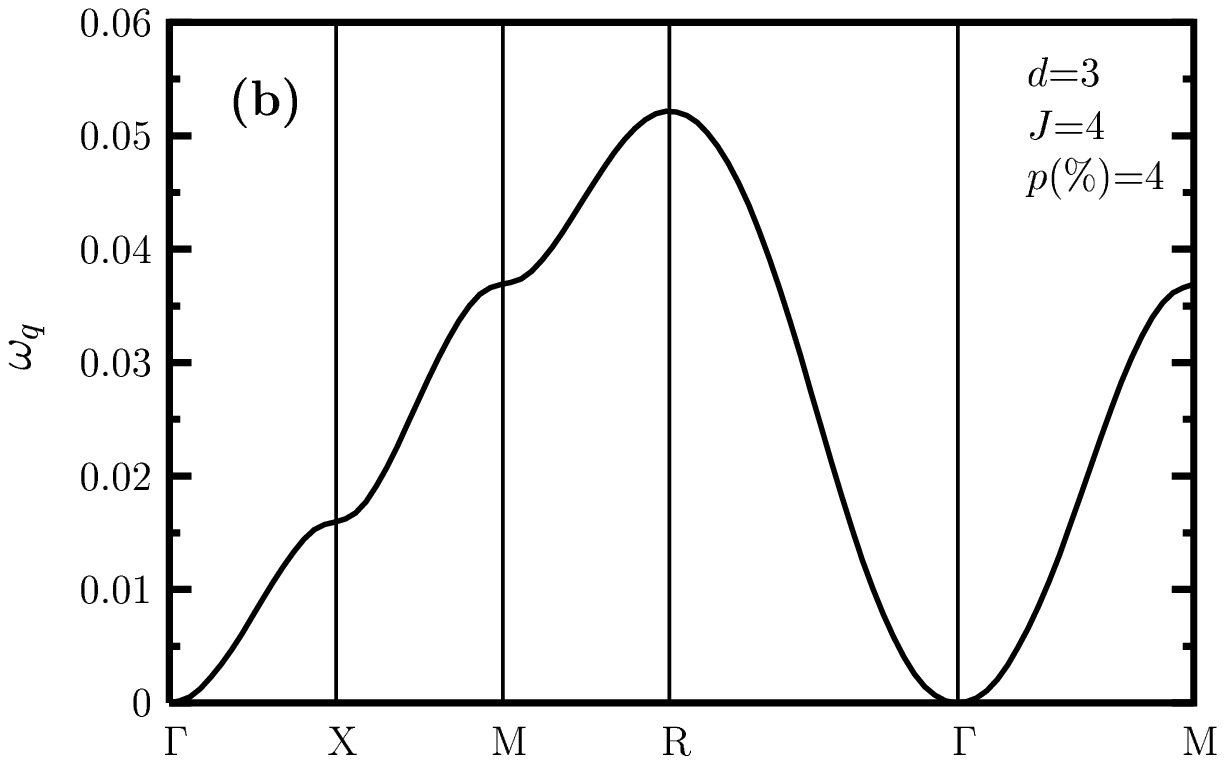,width=140mm}
\vspace{-85mm}
\end{center}
\caption{Spin-wave dispersion along main symmetry directions of the
magnetic Brillouin zone for the (a) 2d ($x$=25\%) and (b) 3d ($x$=12.5\%) cases.}
\end{figure}
The magnon propagator can then be expressed as
\begin{eqnarray}
\chi^{+-}({\bf q},\omega) &=& 
\frac{g^0(\omega)}
{1 + \frac{J^2 S}{2} g^0(\omega) \chi^0({\bf q},\omega)} \nonumber \\
&=&
\frac{1}{\omega-({\cal E}_0 - {\cal E}_{\bf q}(\omega))}
\end{eqnarray}
where the site-diagonal zeroth-order magnon propagator 
\begin{eqnarray}
g^0(\omega)=\frac{1}{\omega - {\cal H}^0 _{\rm boson}} = \frac{1}{\omega - {\cal E}_0} 
\end{eqnarray}
involves the magnon on-site energy ${\cal E}_0 \equiv 
\frac{J}{2} \langle \sigma_I^z \rangle$,
corresponding to the energy cost of a spin deviation,
and ${\cal E}_{\bf q}(\omega)=\frac{J^2S}{2}\chi^0({\bf q},\omega)$ 
is the delocalization energy due to magnon hopping
associated with spin couplings ${\cal J}_{IJ} = J^2 \chi^0_{IJ}(\omega)$.
The energy cost ${\cal E}_0$ of creating a local spin deviation 
is exactly offset by this delocalization-induced energy gain
for $q,\omega=0$, consistent with the Goldstone mode. 
Poles in the magnon propagator yield the magnon-mode energies  by solving 
\begin{eqnarray}
\omega_{\bf q}= {\cal E}_0 - {\cal E}_{\bf q}(\omega_{\bf q}) \; .
\end{eqnarray}
The spin-wave dispersion along the main symmetry directions is shown in Fig. 4 
for the 2d and 3d cases.

\begin{figure}
\begin{center}
\vspace*{-70mm}
\hspace*{-38mm}
\psfig{figure=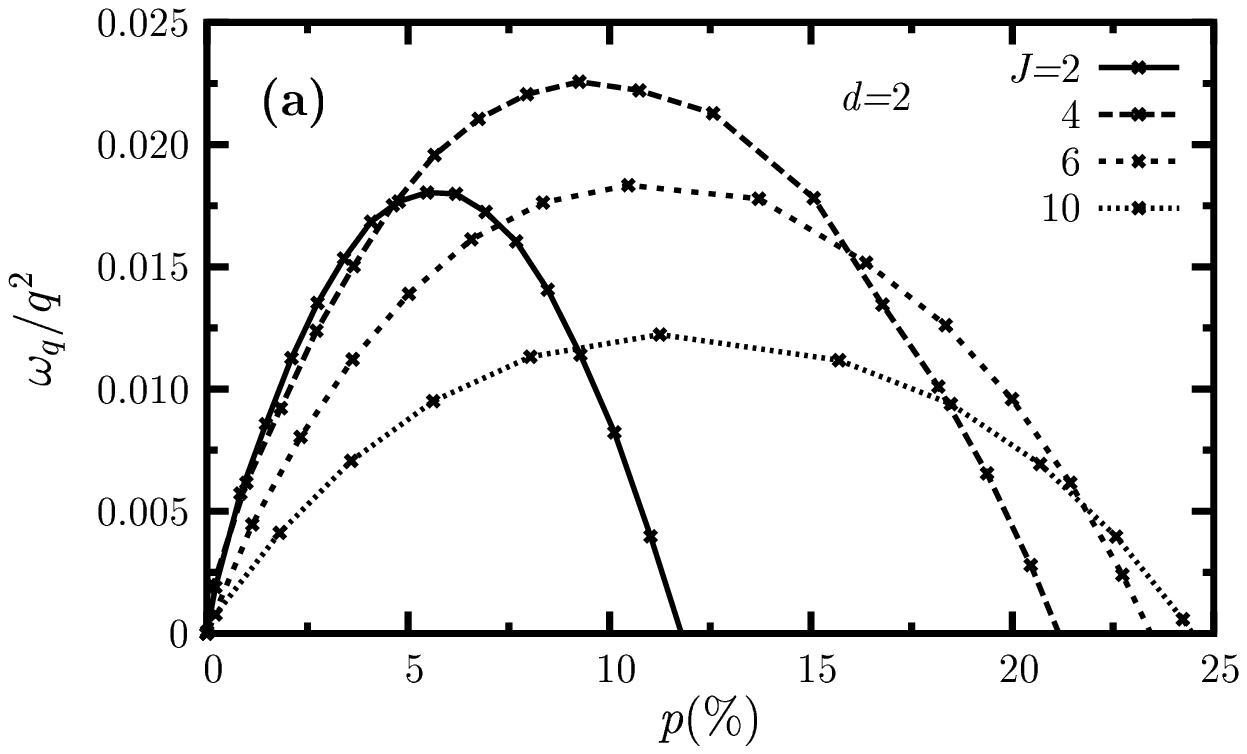,width=140mm}
\vspace{-77mm}
\vspace*{-75mm}
\hspace*{-38mm}
\psfig{figure=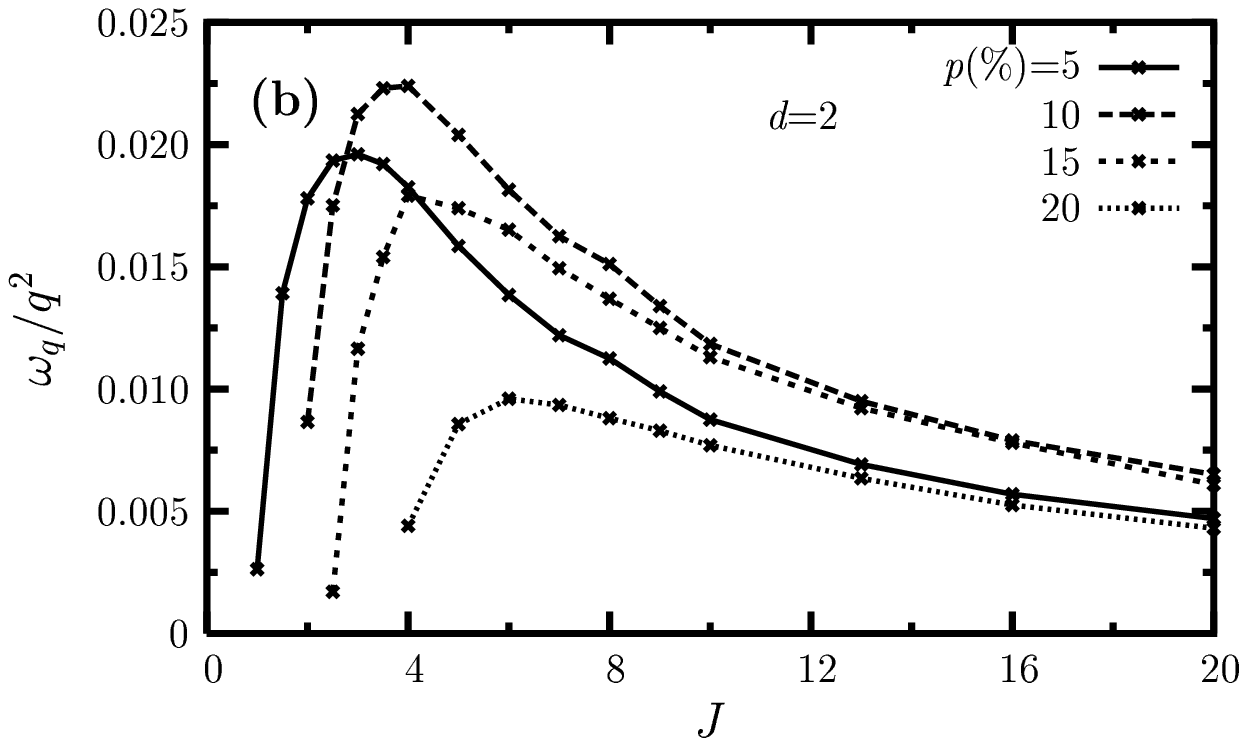,width=140mm}
\vspace{-85mm}
\end{center}
\caption{Variation of spin stiffness with (a) hole concentration $p$ 
and (b) exchange energy $J$ for the 2d case ($x$=25\%).}
\end{figure}

\begin{figure}
\begin{center}
\vspace*{-70mm}
\hspace*{-38mm}
\psfig{figure=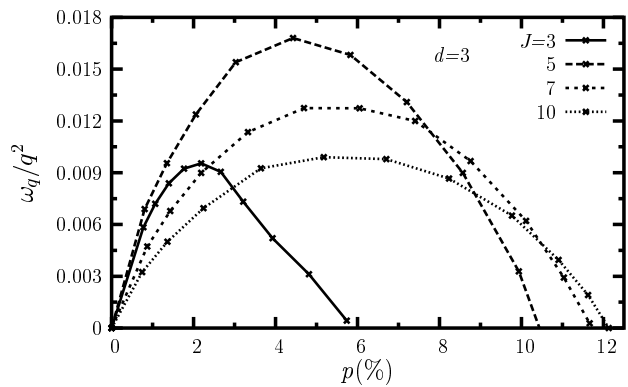,width=140mm}
\vspace{-85mm}
\end{center}
\caption{Variation of spin stiffness with hole concentration $p$
for the 3d case ($x$=12.5\%).}
\end{figure}
\begin{figure}
\begin{center}
\vspace*{-70mm}
\hspace*{-38mm}
\psfig{figure=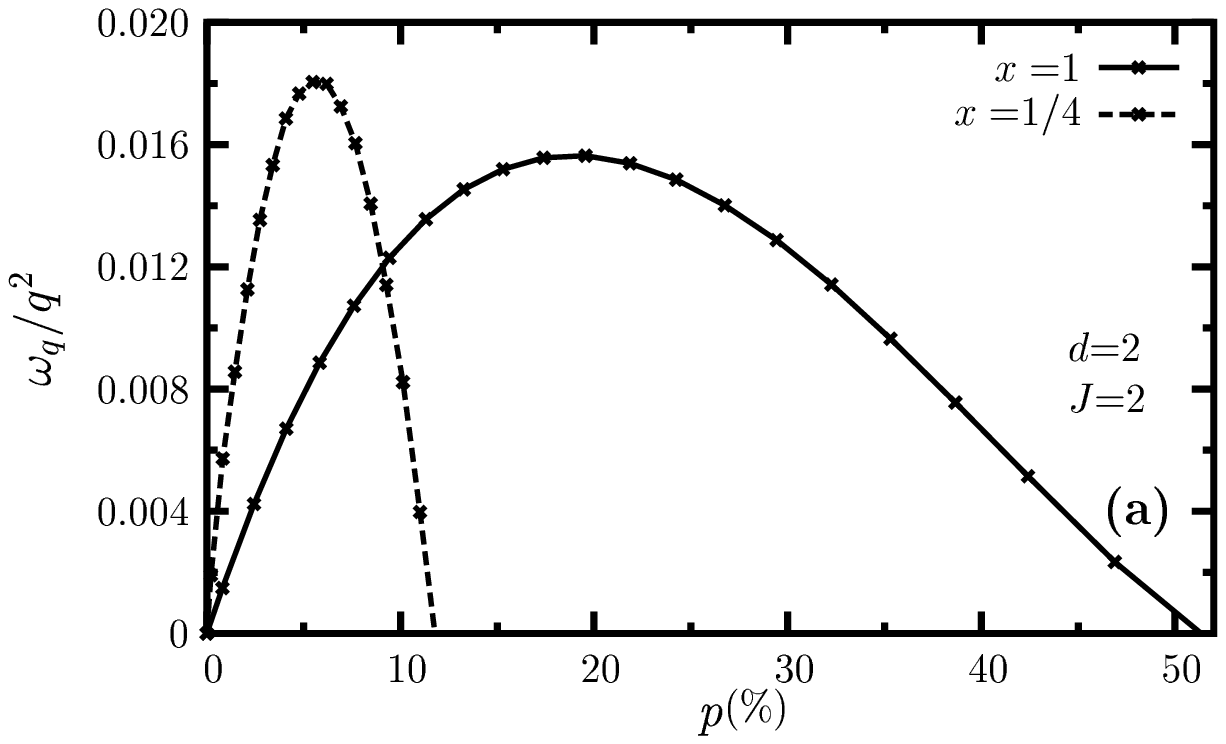,width=140mm}
\vspace{-85mm}
\vspace*{-68mm}
\hspace*{-38mm}
\psfig{figure=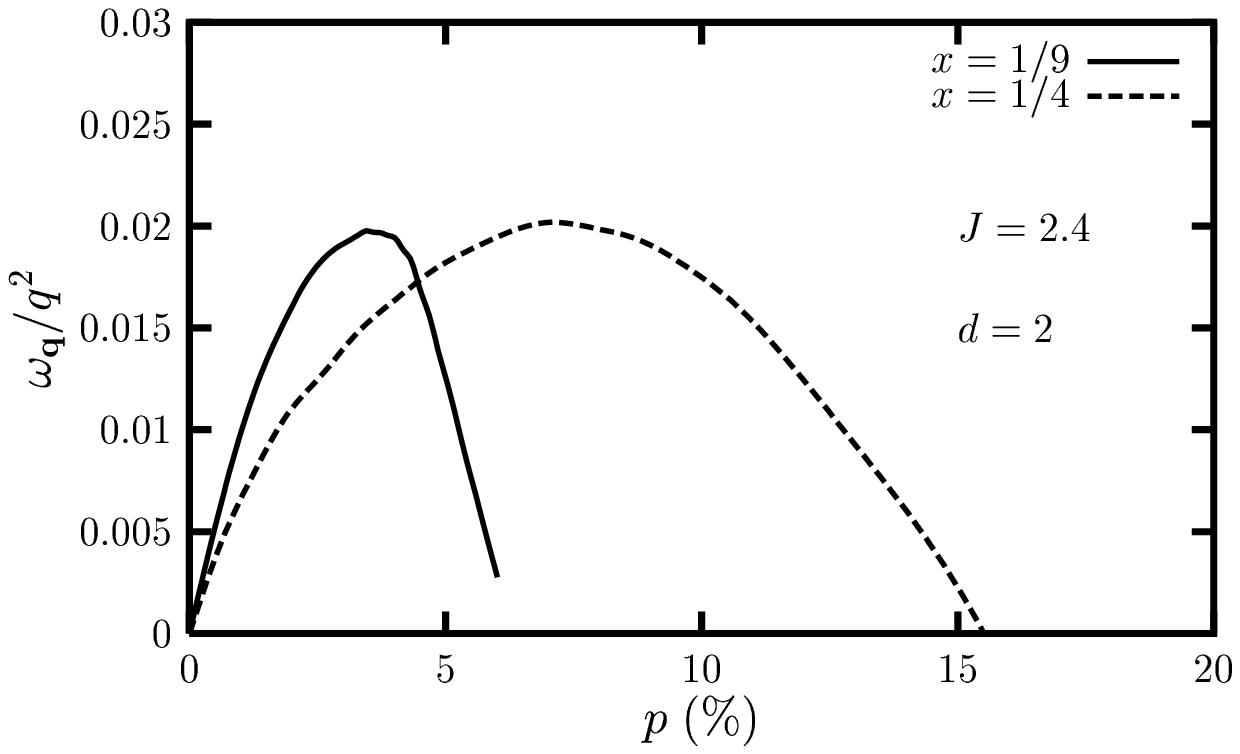,width=140mm}
\vspace{-85mm}
\end{center}
\caption{Comparison of spin stiffness for different impurity spin dilutions ---
(a) $x=1$ and 1/4 and (b) $x=1/4$ and 1/9, 
showing the remarkable enhancement with dilution at low doping.}  
\end{figure}


\section{Spin stiffness}
The spin stiffness $D=\omega_{\bf q}/q^2$ for long-wavelength modes
provides a composite measure of the impurity spin couplings 
and stability of the ferromagnetic state. 
Variation of spin stiffness with doping concentration $p$ and interaction strength $J$
is shown in Figs. 5 and 6 for the 2d and 3d cases.
The spin stiffness exhibits optimization behaviour with respect to both $p$ and $J$,
which can be understood within the generalized RKKY theory in terms of a 
competition between increasing magnitude of
carrier-spin polarization and increasing rapidity of its oscillation.\cite{dms} 
An interesting feature of Figs. 5(a) and 6 
is that due to impurity-host decoupling for higher $J$ values, 
the instability of the ferromagnetic state shifts towards "half-filling" 
($p \rightarrow x$) where AF ordering is favoured.
For a fixed $p$, a minimum $J$ is required to stabilize the ferromagnetic state,
as seen in Fig. 5(b), 
and the $1/J$-type behaviour for large $J$  
is associated with narrowing of the impurity band
due to decreasing effective impurity hopping $t_{\rm eff} \sim (t^2/J)$.

Fig. 7 highlights the remarkable effect of dilution on spin stiffness 
--- dilution actually enhances the spin stiffness at low hole doping,
demonstrating an optimization with respect to impurity concentration as well. 
The above result highlights the key requirements of impurity spin dilution
as well as low hole doping concentration for a robust ferromagnetic state, 
both of which are met in the 
heavily compensated material $\rm Ga_{1-x} Mn_x As$, with $x \sim 5\%$.
It is the reduced particle-hole energy gap in the fermion polarization bubble 
$\chi^0({\bf q},\omega)$ for the diluted case 
which relatively enhances the oscillating carrier-spin polarization,
resulting in greater spin stiffness for low doping and higher sensitivity 
to competing interactions with increasing doping.  
Indeed, the limit of vanishing energy gap (RKKY theory) yields an upper bound to the 
spin stiffness and transition temperature (see Fig. 9).

Fig. 8 shows the enhancement of spin stiffness with dilution at low doping 
for the 3d case. 
Also shown is a comparison of the spin stiffness 
with the VCA result for a uniform impurity field $B=xJS/2$ with
identical lattice-average. 
The VCA result provides a fair approximation in the low doping regime, 
which can be understood in terms of an averaged impurity field 
seen by doped carriers with wavelength longer than the impurity spacing.

\section{Transition Temperature}
Determination of the magnon spectrum in the carrier-induced ferromagnetic state 
allows for an estimation of the Curie temperature $T_c$ in three dimensions.
As the ferromagnetic state is characterized by small spin stiffness due to
competing interactions, the dominant contribution to reduction in magnetization 
is from the thermal excitation of long-wavelength magnon modes. 
Therefore, $T_c$ can be estimated in terms of an equivalent Heisenberg 
model with matching spin stiffness.
For a spin-$S$ Heisenberg ferromagnet with nearest-neighbour interaction $\tilde{J}$ 
on a simple cubic lattice (coordination number $z$=6), the magnon energy 
\begin{eqnarray}
\omega_{\bf q}=z\tilde{J}S(1-\gamma_{\bf q}),
\end{eqnarray}
where $\gamma_{\bf q}$=(cos2$q_x+$cos2$q_y+$cos2$q_z$)/3, corresponding to the 
magnetic lattice spacing 2. 
Indeed, the magnon energy is maximum for $q_x=q_y=q_z=\pi/2$,
as also seen in Fig. 4 (at R).
Considering the small $q$ limit, 
we obtain the spin stiffness $D=\omega_{\bf q}/q^2=4\tilde{J}S=10\tilde{J}$ for $S=5/2$. 

\begin{figure}
\begin{center}
\vspace*{-70mm}
\hspace*{-38mm}
\psfig{figure=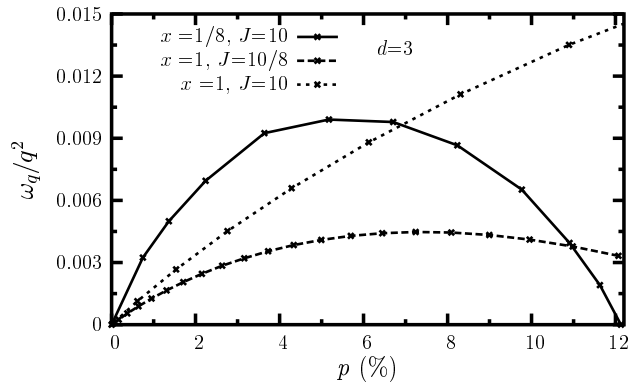,width=140mm}
\vspace{-85mm}
\end{center}
\caption{Comparison of spin stiffness for the 3d case 
with $(x=1/8)$ and without ($x=1$) dilution for same $J$, 
and with the VCA result $(x=1, J=10/8)$.}  
\end{figure}

Within the renormalized spin-fluctuation theory, the transition temperature 
\begin{eqnarray}
T_c^{\rm SF}=\frac{z\tilde{J}S(S+1)}{3}f_{\rm SF}^{-1} = \frac{1}{2} D(S+1)f_{\rm SF}^{-1}
\end{eqnarray}
where the spin-fluctuation factor 
$f_{\rm SF}=\sum_{\bf q}(1-\gamma_{\bf q})^{-1} \approx$ 1.5 for the cubic lattice. 
As $T_c ^{\rm SF} \ll T_c ^{\rm MF}$, 
the impurity mean field is essentially unchanged at low temperature;
therefore we have assumed the zero-temperature value for the spin stiffness.
In the RKKY-VCA approach, the lattice-averaged impurity field 
is assumed to decrease with impurity magnetization and 
vanishes as $T \rightarrow T_c$.\cite{Konig2,Bouzerar}
Considering a bandwidth of the order of that of $\rm GaAs$ ($W=12t \sim 10$ eV), 
we obtain $T_c^{\rm SF}=(35/36)D$ in eV.
The calculated $T_c^{\rm SF}$ (Fig. 9) is in qualitative agreement with 
experimental results for Ga$_{1-x}$Mn$_x$As. 

Figure 9 also provides a comparison with the RKKY result\cite{Dietl1}
\begin{eqnarray}
T_c^{\rm RKKY} = xJ^2\frac{S(S+1)}{12k_{\rm B}} N(E_F),
\end{eqnarray} 
obtained within the mean-field approximation 
from the carrier-induced RKKY spin couplings 
${\cal J}_{IJ} = J^2 \chi^0 _{IJ}$. 
In the limit of low doping concentration ($p \ll 1$),
the host density of states
\begin{eqnarray}
N(E_F) = \frac{1}{2\pi^2} (3\pi^2 p)^{1/3} \; ,
\end{eqnarray}
yielding the characteristic $p^{1/3}$ behaviour of $T_c ^{\rm RKKY}$. 
It is seen in Fig. 9 that the RKKY result drastically over-estimates $T_c$, 
mainly due to neglect of the impurity-field-induced energy gap in the 
fermion polarization bubble, 
and also fails to capture the optimization behaviour with carrier doping,
impurity spin dilution, or interaction strength. 
The MF transition temperature $T_c ^{\rm MF}$, which physically represents
the moment-melting temperature, is an order-of-magnitude still higher.
\begin{figure}
\begin{center}
\vspace*{-70mm}
\hspace*{-38mm}
\psfig{figure=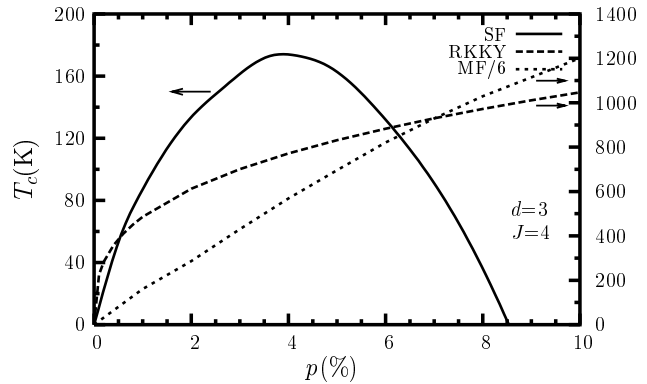,width=140mm}
\vspace{-85mm}
\end{center}
\caption{Comparison of the transition temperature $T_c$ 
obtained from the renormalized spin-fluctuation theory (SF) 
with the RKKY and MF results for the 3d case ($x$=12.5\%).}
\end{figure}
\section{Conclusions}
Magnon excitations in the carrier-induced ferromagnetic
state of the diluted FKLM were studied for several ordered impurity arrangements. 
Use of momentum-space representation allowed for a refined study with respect to variation
of doping concentration, interaction strength, impurity spin dilution, and wave vector.  
The spin stiffness was found to exhibit a characteristic optimization behaviour
with respect to hole doping concentration, interaction strength, as well as 
impurity spin dilution, 
which can be qualitatively understood in terms of a competition between 
increasing magnitude of the carrier-spin polarization and increasing rapidity
of its oscillation.

The exact treatment of impurity spin dilution and interaction strength  
allowed for a detailed comparison with results of existing theories 
such as MF, RKKY, and VCA.
We found that the RKKY result drastically over-estimates $T_c$
due to neglect of the impurity-field-induced Zeeman splitting of the carrier spin bands, 
and also does not capture the optimization behaviour with carrier doping,
impurity spin dilution, or interaction strength. 
The VCA approach involving a uniform lattice-averaged impurity field
was found to provide a fair approximation for spin stiffness in the low doping regime,
but fails to describe the competing interactions at higher doping 
when the hole wavelength becomes comparable to impurity spacing.

The enhancement of spin stiffness with dilution at low doping concentration
due to the reduced impurity-field-induced energy gap in the fermion polarization bubble
highlights the key requirements of impurity spin dilution 
as well as low hole doping concentration for a robust ferromagnetic state
of diluted magnetic semiconductors, both of which are realized in the 
heavily compensated material $\rm Ga_{1-x} Mn_x As$, with $x \sim 5\%$.


\end{document}